\documentclass[reprint,nofootinbib]{revtex4-2}

\usepackage{hyperref}
\usepackage{amsmath}
\usepackage{lipsum}

\usepackage[utf8]{inputenc}
\usepackage[T1]{fontenc}
\usepackage{lmodern}
\usepackage{setspace}
\usepackage{amsfonts}
\usepackage{amssymb}
\usepackage{bm}
\usepackage[dvipsnames]{xcolor}
\usepackage{tabularx}
\usepackage{comment}
\usepackage{mathtools}

\usepackage[]{natbib}

\usepackage{graphicx}
\usepackage{lineno}

\usepackage{csvsimple}
\usepackage{booktabs}
\usepackage{siunitx}

\usepackage{tikz}
\usepackage{pgfplots}
\usepackage{pgfplotstable}
\usepgfplotslibrary{groupplots}
\usetikzlibrary{arrows,arrows.meta,shapes,positioning,3d,datavisualization.formats.functions, pgfplots.groupplots}

\begin{document}

\title{Universality in the fracture of silica glass}

\date{\today}

\author{Somar Shekh Alshabab}
\affiliation{Institute of General Mechanics, RWTH Aachen University, Aachen, Germany}
\author{Bernd Markert}
\affiliation{Institute of General Mechanics, RWTH Aachen University, Aachen, Germany}
\author{Franz Bamer*}
\affiliation{Institute of General Mechanics, RWTH Aachen University, Aachen, Germany}


\begin{abstract}

The presence of universality of avalanches characterizing the inelastic response of disordered materials has the potential to bridge the gap from micro- to macroscale. In this study, we explore the statistics and the scaling behavior of avalanches in the fracture of silica glass on the microscale using molecular mechanics. We introduce a robust method for capturing and quantifying the avalanches, allowing us to perform rigorous statistical analysis, revealing universal power laws associated with critical phenomena. The computed exponents suggest that nanoscale fracture of silica belongs to the same universality class as depinning models. Additionally, the influence of an initial crack is explored, observing deviations from mean-field predictions while maintaining criticality. Furthermore, we investigate the strain-dependent probability density function (PDF), its cutoff function, and the interrelation between the critical exponents. Finally, we unveil distinct scaling behavior for small and large avalanches of the crack growth, shedding light on the underlying fracture mechanisms in silica glass.

\end{abstract}

\maketitle

\section*{Introduction}

Predicting the fracture of amorphous solids is a long-standing research problem. In this context, statistical models have been extensively used to study the breakdown phenomena, since they can address size effects and capture the fluctuations between different samples in real life \cite{herrmann}. 

Since the fracture process is an extremely complex phenomenon that seems to originate from the finest details, the necessity for statistical treatment was recognized early on in the theory of Griffith that takes into consideration the random nature of the defects in the material \cite{griffith}. Later, it was acknowledged that thermally activated fracture could be connected to first-order transition close to the spinodal point \cite{selinger} so that efforts dealt with transition induced by disorder, where the quenched disorder plays the role of temperature \cite{zapperi1999pre}. 

When disordered solids are slowly driven by an external deformation field, their response is characterized by intermittent dynamics in the form of avalanches that increase in intensity before the breakdown of the material \cite{bamer2023, exp_int_dynamics, exp_soc}. These avalanches originate from localized events \cite{stz} and seem to follow power law statistics that are also present in many seemingly different systems beyond fracture and material science \cite{sethna_crackling, fisher_earthquakes, friedman_neurons, biological_evolution}, suggesting a general mechanism behind a more broad phenomenon. Predicting every event related to fracture turns out to be excruciatingly difficult and has been the focus of many studies using different approaches, including structural indicators \cite{structural_indicators}, local shear modulus \cite{patinet_local_yield}, harmonic and anharmonic approximation of the glass' energy \cite{kapteijns_anharmonic, richard_harmonic}, and machine learning \cite{tian_machine_learning}. However, statistical physics can explain the average behavior using simplified models that may capture the macroscopic response of a wide range of systems. These models are generally treated in the mean-field framework \cite{mean_soc} where many of the interaction details of the system are washed away while few relevant properties such as the underlying symmetries play the main role in deciding the leading terms of the expansion of the Landau-Ginzburg free energy \cite{stanley_1987} leading to different scaling relations. 

Similar behavior has been observed for glasses as a major class of disordered solids with widespread applications in various fields due to their high strength and durability. On the nano-scale, (slowly driven) glasses under shear stress experience a steady state phase which has been studied in the framework of self-organized criticality \cite{bak_1987, tang, bak_1996, sun_soc_shear}, while the yielding transition has mostly been studied with depinning models \cite{jie_lin}. However, there is ongoing debate regarding the nature of this transition, with some researchers suggesting that it may belong to a different universality class \cite{budrikis_nc, procaccia}. Moreover, most molecular dynamics (MD) studies were performed on model glasses \cite{oyama_PRE, cyclcic_LJ, driving_rate, yield_xu, yield_gaurav}, or metallic glasses \cite{metallic_yield, demkowicz_yield_silicon}, while only a few have focused on silica glass under shear \cite{PRL_bhaumik} and even less on silica under tensile stress \cite{bonfanti_2018}, where the yielding phase represented by the plastic steady state before failure is quite short-lived. Furthermore, the validity of the mean-field calculations remains under debate \cite{budrikis_2013}, with a wide range of computed critical exponents both numerically and experimentally \cite{sun_soc_shear, exp_slip_avalanches, exp_bmg}. Therefore, a realistic numerical model simulated with MD can be used as a powerful tool to get a better understanding of this complex behavior.

Silica glass which stands out as one of the most widely used materials, is extremely brittle on the nano-scale making it susceptible to catastrophic failure under stress. Thus, its widespread use is severely limited by its cracking properties. Therefore, a better understanding of the failure mechanism can lead to the development of stronger and more fracture-resistant glasses and enables better and more efficient designs.
The deformation behavior of silica is governed by its network topology \cite{bamer2019}, where the breakage of covalent bonds plays an important role in its inelastic response \cite{bamer2021, ebrahem2020}. Furthermore, experimental studies have shown that cracks in silica glass grow in a sequential bond rapture without plastic deformation near the crack tip \cite{lawn1980}, while some other experimental \cite{Celarie2003} and molecular dynamic studies \cite{Rountree2007} claim that it breaks through the coalescence of nanoscale cavities and that plastic flow governs the cracking process \cite{Barthel2020}.  Furthermore, it has been shown that, the crack geometry exhibits the universal values of the crack roughness \cite{Rountree2007}, providing further indication of the universality of the fracture process.

In this paper, we use molecular dynamics to study the fracture process of silica glass on the nano-scale under tensile stress, using the athermal quasi-static (AQS) deformation protocol for a wide range of system sizes including pre-notched and un-notched systems. A new method is proposed for detecting and measuring the avalanches in the stress and energy of the system. Their statistics are analyzed and compared, showing that they follow power laws suggesting critical behavior. The critical exponents are determined by employing direct fitting and finite-size scaling (FSS) techniques. Additionally, we investigate the dependence of the PDF on the strain range, explore its cutoff function, and examine the relationship between the critical exponents. Finally, the crack growth process is captured, and the statistics of the corresponding avalanches are analyzed. 

\section{Methods}

\subsection{Sample generation}
We investigate the fracture of molecular systems using classical molecular mechanics. Four different system sizes of bulk silica were simulated, containing 6100, 12150, 24624, and 48000 atoms. The largest system has the dimensions $88 \times 56 \times 140$ \AA\ and a density of $2.2$ g/cm\textsuperscript{3}. The other systems maintain approximately the same proportions and density. Each system size was simulated by heating an ensemble to $8000$ K. Subsequently, the temperature is kept constant using the NVT ensemble for $250$ ps to ensure that, firstly, the samples reached a state of thermal equilibrium and, secondly, they have no memory of their initial configuration. The system is then rapidly quenched to a temperature of $0.01$ K at the rate of $10$ K/ps using the NVT ensemble. Finally, the samples are allowed to relax to a pressure-free state by iteratively altering the simulation box and minimizing the potential energy. We used a potential based on the Born-Mayer-Huggins potential, first proposed by Matsui \cite{matsui}, and re-parameterized by Jakse et al.~\cite{jakse}. The potential between two atoms of type $i$ (Si) $j$ (O) is written as:

\begin{equation}
 U_{ij}(r_{ij}) = \frac{q_i q_j}{4\pi\varepsilon_0r_{ij}} + A_{ij}\exp(\frac{\sigma_{ij} - r_{ij}}{\rho_{ij}}) - \frac{C_{ij}}{r_{ij}^6}  \;,
 \label{eq:Poten}
\end{equation}

where $r_{ij}$ is the inter-atomic distance, $q_i$ refers to the effective charge, as well as $A_{ij}$ and $C_{ij}$ are parameters taken from Jakse et al.~\cite{jakse}.
Equation (\ref{eq:Poten}) includes three terms: the Coulombic interaction term, short-range repulsive term (also known as the Born term), and the Van der Waals interaction term. The cut-off distance was set to $ 10.17$ \AA.

\subsection{Mechanical Simulation}\label{time_velocity}

To investigate crack growth, an initial notch was introduced to the center of the samples by removing atoms from a cylindrical volume with an elliptical cross-section of $2.5\times5$ \AA\ in the $xz$-plane which spans across the entire sample in the $y$-direction.
The size of the initial notch was chosen to be as small as possible while making sure that it constitutes a critical size for initiating failure. The samples are loaded uni-axially in the $x$-direction at $0$ K using the athermal quasistatic deformation protocol \cite{maloney} with a strain step size of $1\times10^{-4}$. Notably, the step size was chosen small enough to make sure that it captures all the avalanches occurring during deformation. The average time between the avalanches (weighting time) was measured for the biggest sample to be $\sim 2\times10^{-3}$.

Following the described methodology, $50$ samples of each system size were generated and loaded until failure. A total number of 23025 avalanche events were analyzed for the investigation of the energy drops, stress drops, and 11253 events for the crack growth. 

\subsection{Void detection}\label{Voids}

To study the crack growth during mechanical testing, we used an external library \cite{voloshin, sastry} with which both the volume and the surface of the voids can be computed. In this paper, the voids were defined as the available space through which a hard sphere can move by assigning a spherical volume around every atom in the system.
A diameter of $2$ \AA\ was chosen for both the exclusion spheres around each atom and the moving hard-sphere taking into account the covalent bond distance between silicon and oxygen atoms.

\subsection{Scaling}

Three different quantities were chosen for the quantification of avalanche events: the intermittent drops in the virial stress during deformation $\Delta\sigma$, the drop in the potential energy $\Delta \Pi$, and the discrete jumps in the volume starting from the initial notch volume $\Delta V$ which identifies the crack growth.

\begin{figure}
    \centering
    \includegraphics{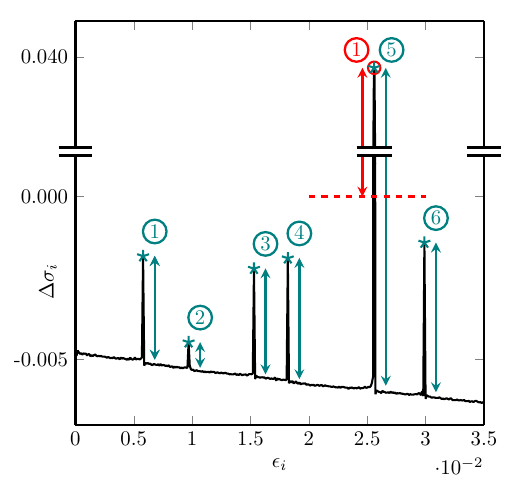}
    \caption{Illustrative plot of the avalanche measurements. The black line represents the stress difference between two consecutive steps $\Delta \sigma_{i} = \sigma_{i+1} - \sigma_{i}$ as a function of the strain $\varepsilon_i$. The green color shows the detected avalanches if one considers a wider range of fluctuations, and the red color shows the detected avalanches if one defines avalanches as $\Delta \sigma_i > 0$. The arrows represent the avalanche magnitude of each method. The inclination in the black line shows how the elastic response of the material changes with increasing strain.}
    \label{fig:measurements} 
\end{figure}

Since the behavior in the response of the system changes with increasing strain, the conventional way of measuring avalanches, which is simply to evaluate the difference in the observed response magnitudes, such as stress or potential energy at two subsequent strain steps, leads to crucial inaccuracies for the present examples. On the one hand, it would lead to an overestimation in the avalanche size at the initial elastic regime, and, on the other hand, it would lead to an underestimation in the regime closer to the critical strain. In the case of steady flow as in a sheared glass, one could overcome this problem by correcting the drop according to the expected value in the case of a continued elastic behavior by using the elastic constant as follows $\Delta\sigma = \sigma_i - \sigma_{i+1} + E \Delta\varepsilon $ \cite{lemaitre2007}. However, this is not straight forward in our case since the elastic constant changes significantly with increasing strain.
Therefore, we propose an alternative method for measuring the avalanches wherein events are detected in terms of fluctuations in the studied quantity rather than by simply identifying the drops. To this end, the avalanches are identified as discontinuities in the change of the energy ($\Pi_i - \Pi_{i+1}$) or stress ($\sigma_i - \sigma_{i+1}$). Then a running average is constructed at the time steps without events, and interpolated at the time of the events, which would correspond to a running average curve of the elastic constant $E$. Finally, the size of the avalanches is computed as the difference between the peaks and the running average curve. An illustration of the approach is shown in Figure \ref{fig:measurements}.

Taking into account relevant literature published in fracture mechanics, we expect the avalanche statistics to follow a power law with a system size-dependent cutoff, cf.~\cite{zapperi1999pre, budrikis_nc}. Based on this assumption, we initially computed the critical exponent from the simulations by direct fitting. However, it is known that fitting with the moments of the PDFs by assuming finite-size scaling yields better estimates, particularly when the statistical sample size is inadequate for determining the cutoff tail with precision \cite{zapperi_scaling}. Therefore, we carried out finite-size scaling analysis by investigating the moments of the distribution function from which we estimated the critical exponents. The validity of this assumption can be assessed by evaluating the quality of the resulting data collapse.

If we assume that the integrated PDF over a certain strain domain scales as follows:

\begin{equation}
    P(m) \sim N^{-\beta}f(\frac{m}{N^{\beta/\tau}}) \; ,
\end{equation}

where $N$ is the system size. Then, one expects the first moment of the PDF to scale as follows:
\begin{equation}
    \langle m^a \rangle \sim N^{\frac{\beta}{\tau}(a + 1 - \tau)} \sim N^{\alpha(a)} \; .
\end{equation}

This equation is only valid for values of $a+1-\tau > 0$, and it reads:
\begin{equation}
    \alpha(a) = \frac{\beta}{\tau}(a-\tau + 1),\, \frac{\partial\alpha(a)}{\partial a} = \frac{\beta}{\tau} \; .
\label{eq:fss_alpha}
\end{equation}

Therefore, the exponents $\tau$ and $\beta$ can be computed from the log-log plot of $N - \langle m^{\alpha} \rangle$, for different values of $\alpha$, by following the method proposed by Chessa et al.~\cite{zapperi_scaling}.

To study the cut-off tail for the PDF, we rely on the scaling relations derived from the fiber bundle model with the global load-sharing rule, which corresponds to the mean-field calculation for the fracture transition due to the infinite interaction length between the single fibers. Accordingly, we expect the PDF of the avalanche size to scale as follows:

\begin{equation}
P(m, \varepsilon) \sim m^{-\tau}e^{-m(\varepsilon_c-\varepsilon)^{\kappa}} \;.
\label{eq:fbm_avalanches}
\end{equation}

Here, $m$ is the avalanche size, the strain $\varepsilon$ plays the role of the external force field $f$, and $\varepsilon_c$ is the critical strain. This relation is expected to be valid for small bin sizes; however, in our case, a much larger statistical sample size is needed, which poses a challenge. Therefore, we adopt a strategy of integrating the PDF over different strain regimes. By integrating from $\varepsilon = 0$ to $\varepsilon = \varepsilon_c$, it would take the following form:

\begin{equation}
    P(m) \sim  \frac{1}{\kappa} m^{-\tau'} \gamma(1/\kappa, m\varepsilon_c^k) \;,
\end{equation}

where $\tau' = \tau+1/\kappa$ and $\gamma$ is the lower incomplete gamma function. In the case of arbitrary bin sizes $\varepsilon_1 - \varepsilon_2$ it is written as:

\begin{equation}
    \begin{split}
    P(m) \sim  \frac{1}{\kappa} m^{-\tau'} \left( \gamma(1/\kappa, m(\varepsilon_c-\varepsilon_2)^{\kappa}) - \right.\\ \left. \gamma(1/\kappa, m(\varepsilon_c-\varepsilon_1)^{\kappa}) \right) \;.
\end{split}
\end{equation}

Finally, for any strain regime from 0 to $\varepsilon$, it is written as:

\begin{equation}
    \begin{split}
    P(m, \varepsilon) \sim  m^{-\tau'} \left( \gamma(1/\kappa, m(\varepsilon_c-\varepsilon)^{\kappa}) - \right. \\
    \left. \gamma(1/\kappa, m(\varepsilon_c)^{\kappa}) \right) \;.
    \end{split}
    \label{eq:pdf_integrated_0_ec}
\end{equation}

Additionally, one may analyze the moments of the PDFs for different strain regimes, starting from Equation (\ref{eq:fbm_avalanches}). The average avalanche size would diverge as the system approaches the critical strain, according to the following scaling relation:

\begin{equation}
    \langle m^a \rangle \sim (\varepsilon_c-\varepsilon)^{\kappa(\tau-a-1)}\gamma(a+1, Am_c(\varepsilon_c-\varepsilon)^{\kappa})
    \label{eq:avg_avalanche_per_strain} \;,
\end{equation}

where $m_c$ is the critical avalanche size. We assume that it is related to the linear dimension $m_c \sim L^{d_f}$ or $m_c \sim N^{d_f/3}$ where $d_f$ is the fractal dimension, or that $m_c \sim m_{max}$. For the FBM, one expects the critical exponents to be $\tau = 1.5$ and $\kappa = 1.0$ \cite{hansen}.

We assume that pre-existing defects in the material interact through long-range elastic fields \cite{focks}. This could lead to mean-field behavior even in dimensions lower than the lower critical dimension \cite{unger}. One may study the extent of the elastic fields by analyzing the load transfer in the system which is expected to decay exponentially as $1/r^{\gamma}$, where $r$ is the interaction radius. In the fiber bundle model, the interaction range is infinite, and, therefore, the calculations are thought to correspond to the mean-field limit \cite{alava}, while in the ones incorporating the local load sharing rule, $\gamma$ defines the system response to the fiber breaking events, where $\gamma = 2$ forms the cross-over between local and global behavior. The complete derivations of the aforementioned equations are presented in the appendix.

\section*{Results and discussion}
\begin{figure*}
    \begin{tikzpicture}
        \node at (0.0,5.5){\textbf{(a)}};
        \node at (0,-0){
            \begin{tikzpicture}
                \node[inner sep=0pt] at (0,0) {
                \includegraphics[width=6.5cm, origin=lb, trim = 47 38 56.5 39.5, clip]{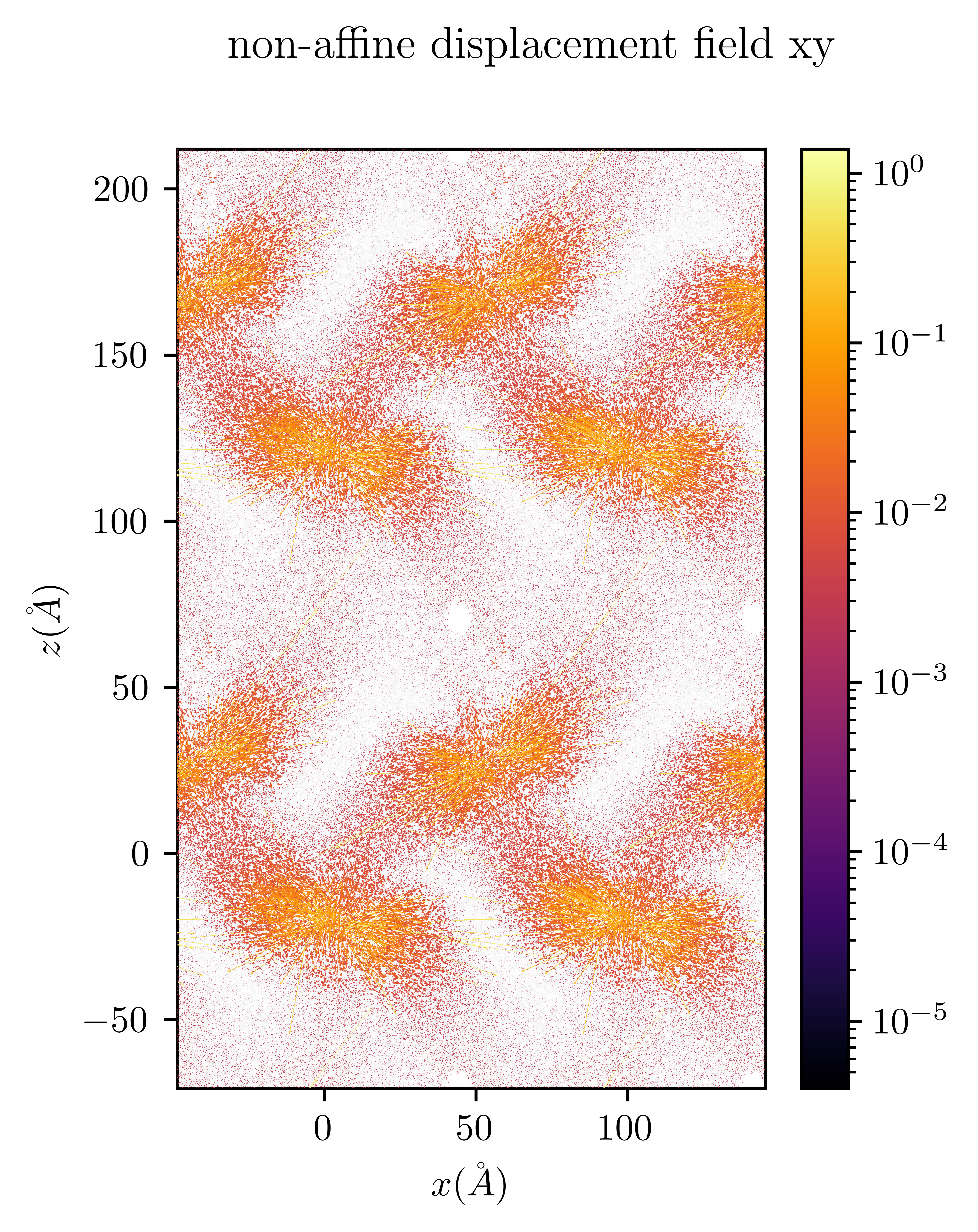}};
            \end{tikzpicture}        
            };
        \node at (8.0,5.5){\textbf{(b)}};
        \node at (8,-0){
            \begin{tikzpicture}
                \node[inner sep=0pt] at (0,0) {
                \includegraphics[width=6.5cm, origin=lb, trim = 111 72 101 65, clip]{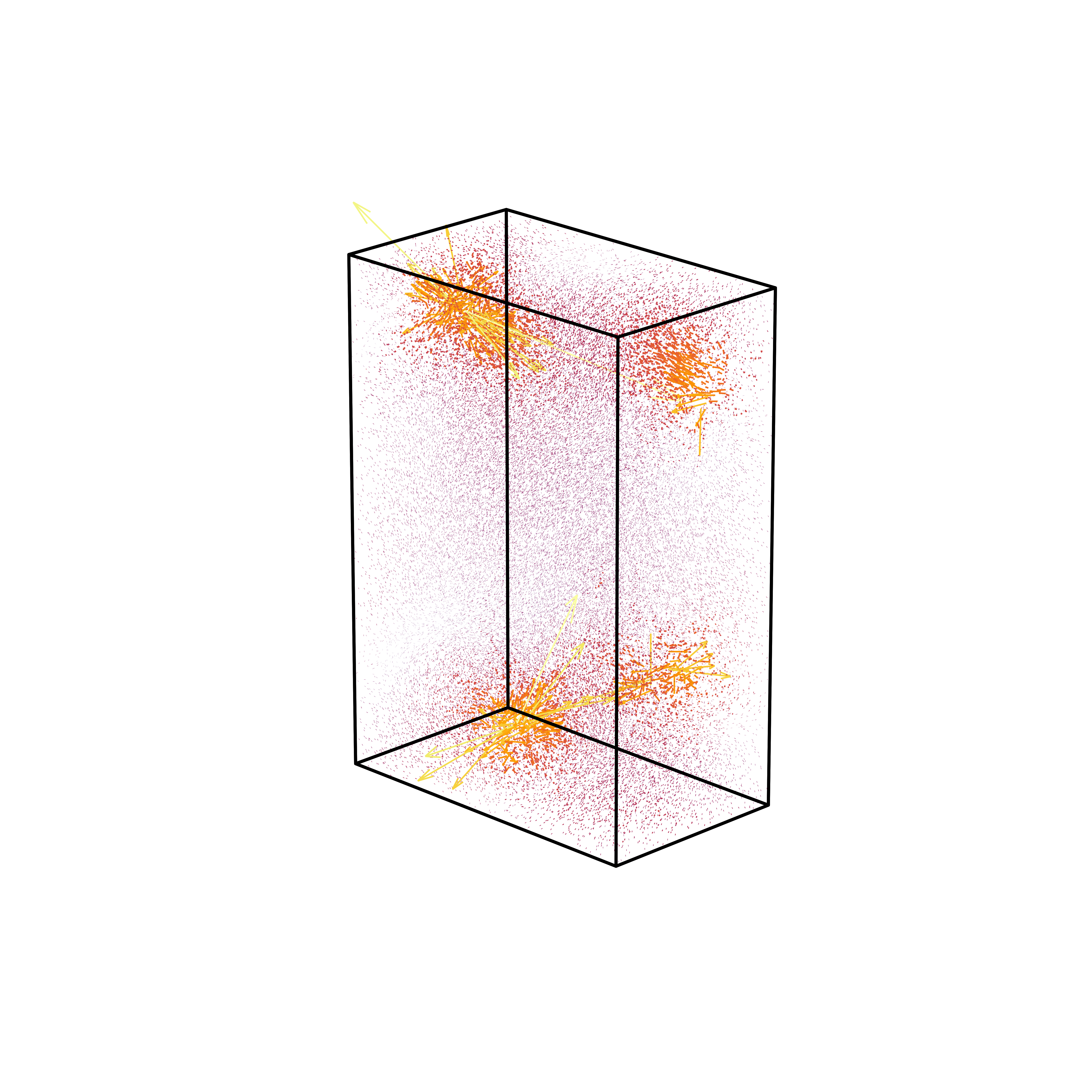}};
            \end{tikzpicture}     
            }; 
        \node at (0.0,-5.75){\textbf{(c)}};
        \node at (0,-11.5){
            \begin{tikzpicture}
                \node[inner sep=0pt] at (0,0) {
                \includegraphics[width=6.5cm, origin=lb, trim = 47 38 56.5 39.5, clip]{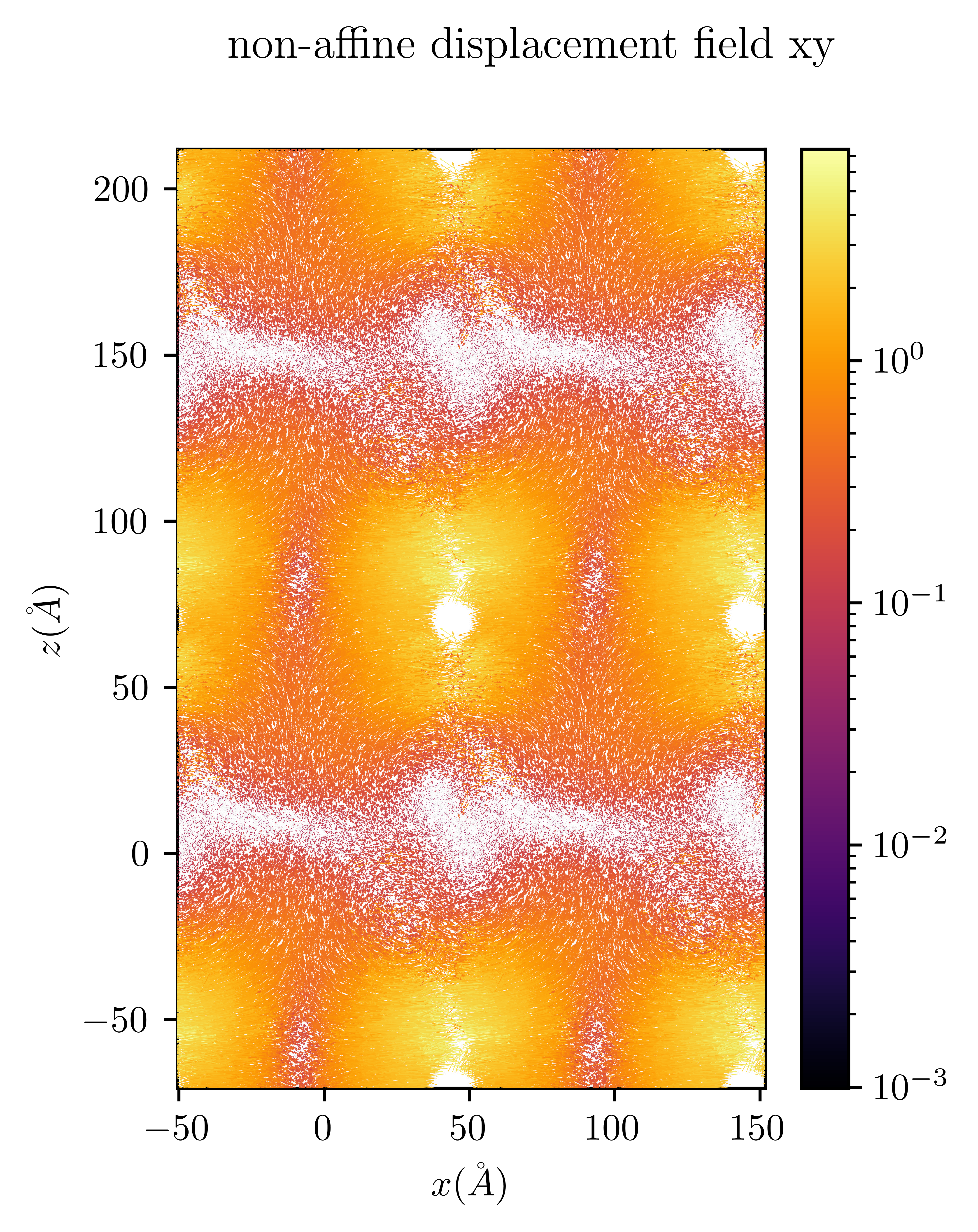}};
            \end{tikzpicture}        
            };
        \node at (8.0,-5.75){\textbf{(d)}};
        \node at (8,-11.5){
            \begin{tikzpicture}
                \node[inner sep=0pt] at (0,0) {
                \includegraphics[width=6.5cm, origin=lb, trim = 75 18 63 17, clip]{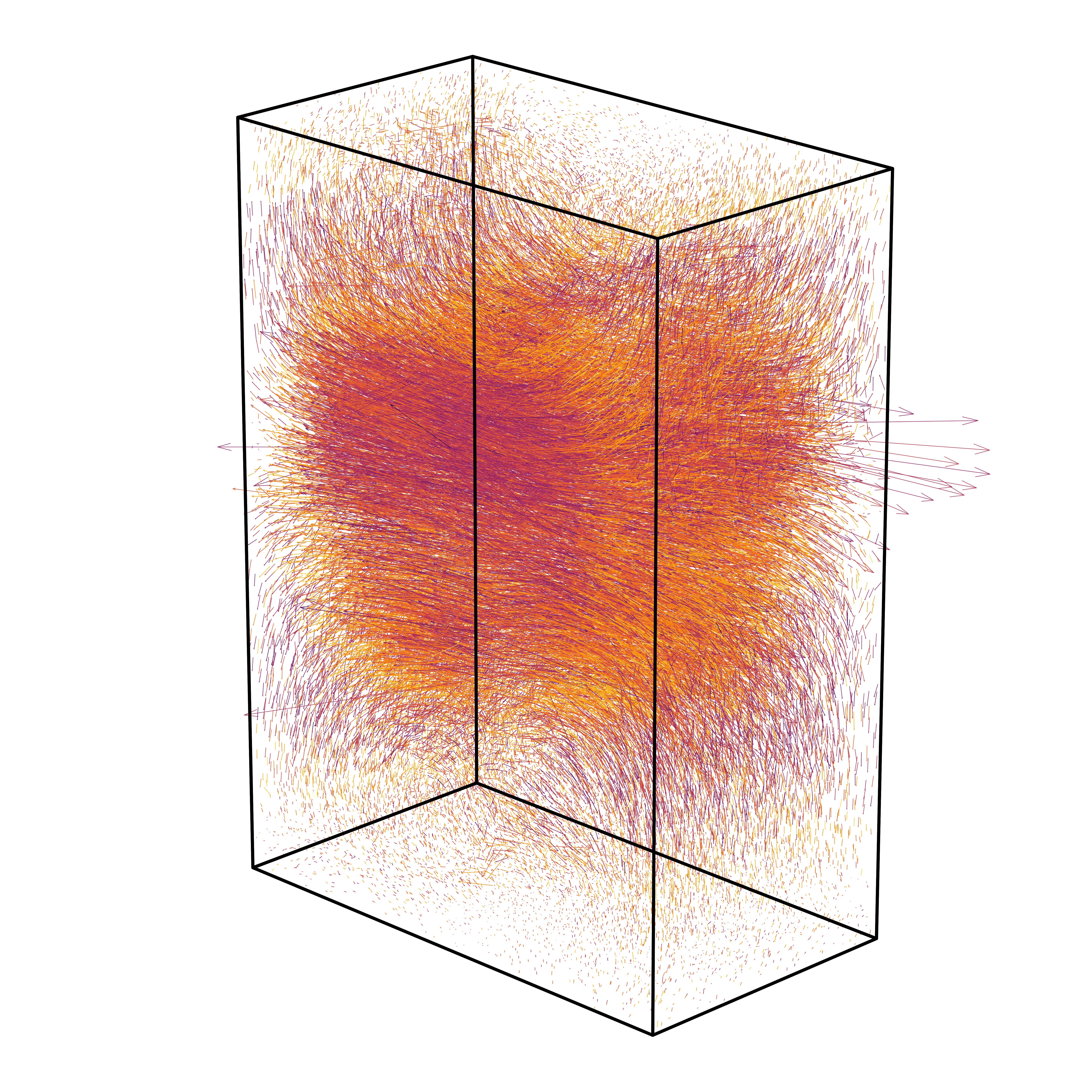}};
            \end{tikzpicture}        
            };            
    \end{tikzpicture}
    \caption{Non-affine displacement field for two events in the system with 48000 atoms; (a) a small event in the elastic strain regime projected on the $xz$-plane; (b) the same event in 3d; (c) the largest event during loading projected on the $xz$-plane, leading to the propagation of the initial crack; (d) the same event in 3d. The arrows and colors are normalized and scaled logarithmically. Hence, the events are more localized than they appear in the plots. Additionally, the plots are replicated in both the $x$- and $z$-direction in (a) and (c) to demonstrate the periodic boundary conditions.}
    \label{fig:non_affine_disp_field} 
\end{figure*}
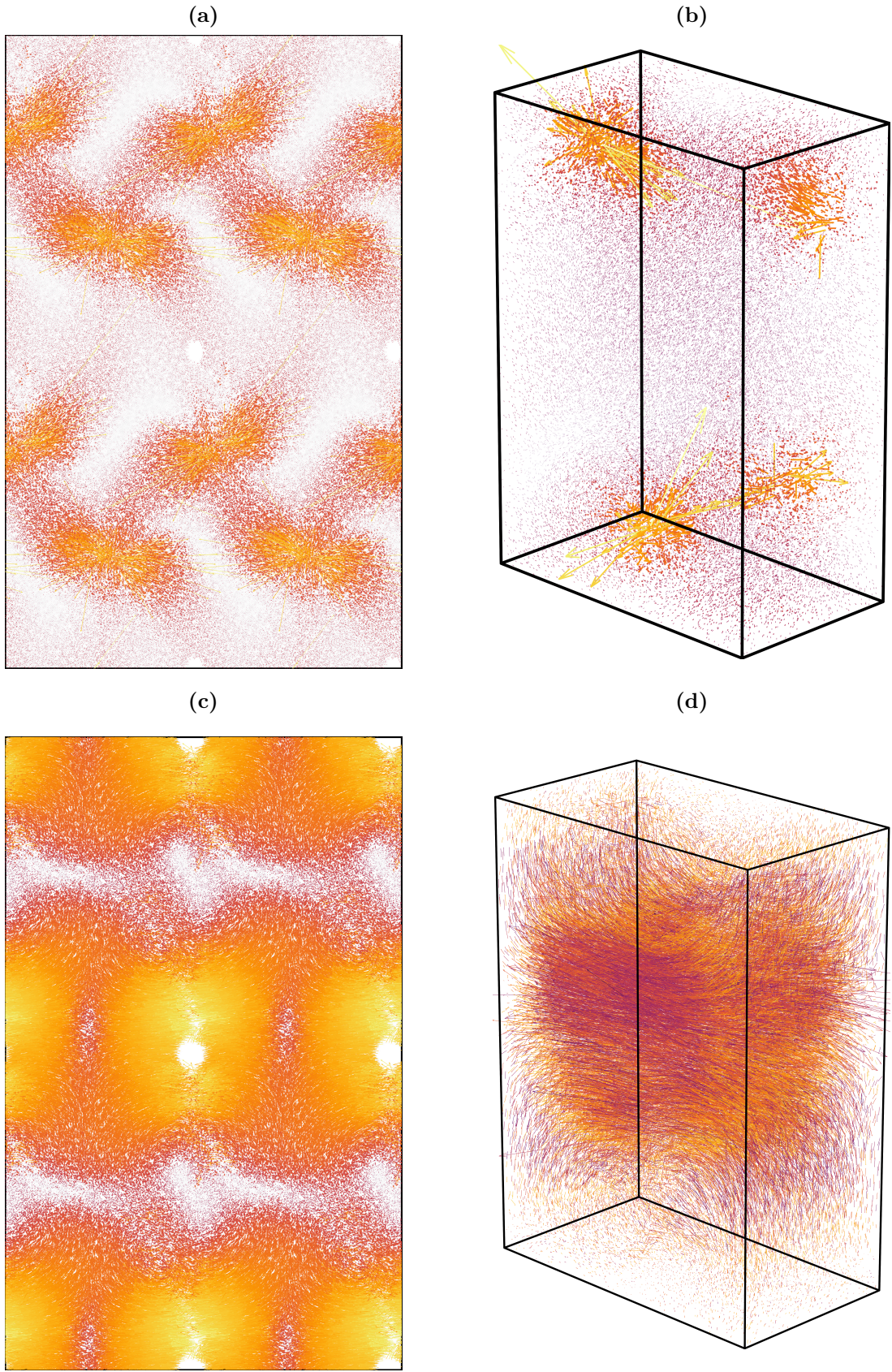

At low strains, the material exhibits nearly linear behavior, and only a relatively small number of plastic events was detected. These events increase in size and frequency with increasing strain until failure. All the simulations reveal brittle behavior. However, the ductility increases with decreasing system size due to size effects. Furthermore, a short steady state before failure was observed as expected from rather brittle materials. Since the presence of a long-lived absorbing phase represented by the steady state is an important criterion for self-organized criticality \cite{zapperi_prl_1997, dickman} one may argue that classifying the fracture of silica in the framework of self-organized criticality may not be feasible. 

The non-affine displacement fields for two avalanche events are plotted in Figure \ref{fig:non_affine_disp_field}. A small event is plotted in \ref{fig:non_affine_disp_field} (a) and \ref{fig:non_affine_disp_field} (b) where it can be seen that the event consists of two sub-events interacting with each other through the elastic fields. However, neither of them is at the edge of the initial notch. This observation highlights the role of the quenched disorder of the system whereby not all events result in the propagation of the crack. Furthermore, the events shown take dipolar shapes with two distinct inclinations relative to the loading axis. It is worth noting that the dipolar shape was predominant in our simulations, in contrast to events in shear loading regimes, which are characterized by quadrupolar shapes \cite{bamer2020}. This difference in shape is known to impact the waiting time between avalanches \cite{lerner}. In Figure \ref{fig:non_affine_disp_field} (c) and \ref{fig:non_affine_disp_field} (d) a catastrophic event is plotted whose magnitude extends over the entire system and leads to failure of the system. It is worth noting the power laws derived from the mean field of depinning models are expected to fail to capture the statistics of the large system-spanning events \cite{dahmen2009}. Moreover, the catastrophic events in the FBM are known to follow a Gaussian distribution \cite{alava}.

\begin{figure*}
    \centering
    \includegraphics{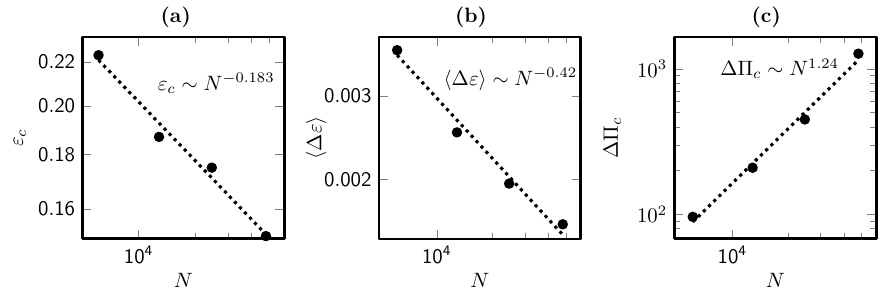}
    \caption{dependence of the critical strain (a), the average waiting time between the energy avalanches (b), and the critical avalanche size on the system size.}
    \label{fig:system_size_dependanc} 
\end{figure*}

The average strain interval between the avalanches decreases exponentially with the increasing system size as $\Delta \varepsilon \sim N^{-\chi}$ with $\chi \approx 0.4234$. Similarly, the critical strain is defined as the strain at which the average avalanche size diverges scales as $\varepsilon_c \sim N^{-B}$ with $B \approx 0.1025$. These results are plotted in Figure \ref{fig:system_size_dependanc}.

By assuming that the average maximum avalanche from each system, corresponds to the critical avalanche size, one can compute the fractal dimension from its dependence on the system size. The results are plotted in Figure \ref{fig:system_size_dependanc} (a). One can see from the slope of the fitted line that ($d_f/3 \approx 1.24$ and, therefore,) $d_f \approx 3.72$. This is much larger than the typical values reported from statistical fracture models, which usually fall in the range between $1.0$ and $1.25$ \cite{alava}. 

\subsection{Avalanches}

\begin{figure*}
    \centering
    \includegraphics{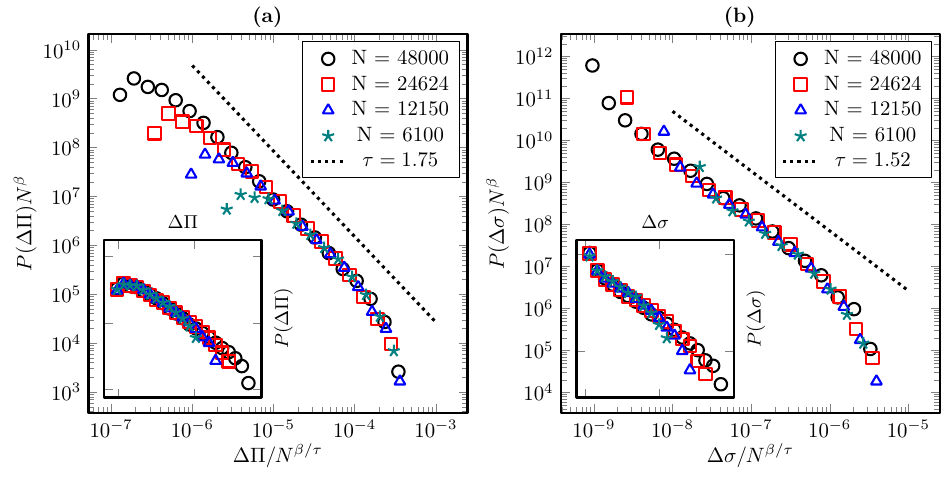}
    \caption{Scaled probability distribution function for the energy (a)and stress (b) avalanches. Inset: the unscaled PDF. The data collapse indicates a good fit for the computed exponents from FSS. The avalanches are integrated over the strain regime until the critical strain. The black line represents a  power law $m^{-\tau'}$ where the critical exponent $\tau'$ has been computed from the first moment analysis of the distribution function.}
    \label{fig:energy_avalanche_pdf_collapsed} 
\end{figure*}

Our analyses confirm the occurrence of scale-invariant avalanches. In Figure \ref{fig:energy_avalanche_pdf_collapsed}, the probability distribution function $P(\Delta\Pi)$, taking into account the strain domain from zero until the critical strain, is plotted. The avalanches follow a power law with cutoffs at both ends of the PDF. The upper cutoff is exponential and scales with the system size. We find through the direct fitting of the sample containing $48000$ atoms that the critical exponent takes a value $\tau'_{\Delta \Pi} \approx 1.75$. However, to confirm this value, we carry finite size analyses by assuming a scaling function in the form $P(\Delta \Pi) \sim N^{-\beta}f(\Delta \Pi/N^{-\beta/\tau'})$. By analyzing the first moment of the PDF, we evaluated $\beta_{\Delta \Pi}/\tau'_{\Delta \Pi} = 1.47$ from the derivative $\partial{\alpha(a)}/\partial{a}$, which was computed from the slope of the logarithmic plot of $N - \langle m^{\alpha} \rangle$. Subsequently, $\tau'_{\Delta \Pi} = 1.75$ was evaluated from the relation $\alpha(a) = \beta/\tau'(a - \tau' + 1)$, so that $\beta_{\Delta \Pi} = 2.57$. Finally, to validate our calculations, we collapse the data with the hypothesized function and the computed exponents in Figure \ref{fig:energy_avalanche_pdf_collapsed}. As shown in this figure, the collapse is satisfactory since the data points of the PDFs of the different system sizes fall into one curve.

\begin{figure}
    \centering
    \includegraphics{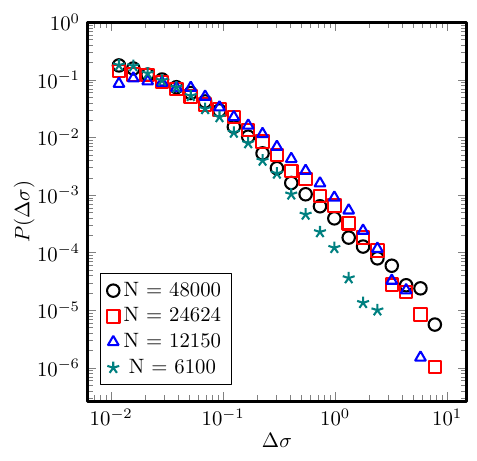}
    \caption{Probability distribution function for the stress avalanches when the measurement is taken as the simple stress drop without stiffness-based corrections. This leads to deviation from the universality of the PDFs of the different systems.}
    \label{fig:stress_drops_pdf_logspace} 
\end{figure}

\begin{figure}
    \centering\includegraphics{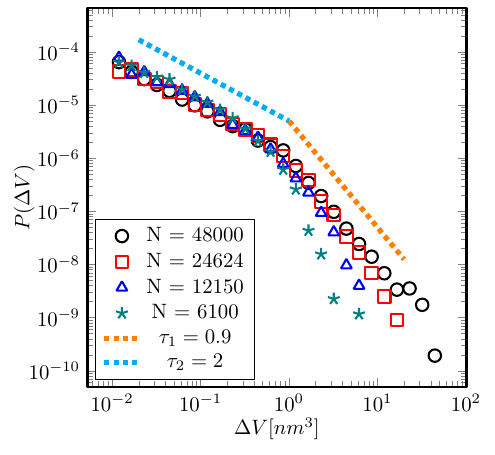}
    \caption{Probability distribution function for the volume avalanches. The measurements have been integrated over the whole loading regime. i.e. until complete failure of the samples. The distribution is not well approximated by a simple power law and instead appears to follow a double power law.}
    \label{fig:volume_avalanche_pdf_logspace} 
\end{figure}

Similarly, the probability distribution function of the stress drops $P(\Delta\sigma)$ is plotted in Figure \ref{fig:energy_avalanche_pdf_collapsed} (b). A power law is observed with cutoffs at both ends, with the upper cutoff being dependent on the system size. We compute $\tau'_{\Delta \sigma} \approx 1.50$ through direct fitting of the biggest system. Subsequently, we computed through FSS $\tau'_{\Delta \sigma} = 1.52$ and $\beta_{\Delta \sigma} = 2.52$, following the same procedure as for the energy avalanches. For comparison, if the avalanche measurement is considered as a simple difference in stress before and after the event, the statistics show a strong bias in the power, contradicting the criticality assumption of the avalanches, as shown in Figure \ref{fig:stress_drops_pdf_logspace}.
The fractal dimension can be estimated also from FSS since we know that the value $\beta/\kappa$ should correspond asymptotically to $d_f/d$, where $d = 3$ is the spatial dimensions of the system. Accordingly, we find that the fractal dimension computed from energy avalanches $d_f\approx 4.4$ confirming that the critical dimension in our systems is much larger than the expected range.
Equally, we demonstrate that crack growth occurs in avalanches that also follow a power law. However, the avalanches appear to deviate from a simple power law and exhibit a multi-fractal distribution, in particular, a double power law. This can be seen in Figure \ref{fig:volume_avalanche_pdf_logspace}, where the probability density function $P(\Delta V)$ is presented. Direct fitting of critical exponents for the biggest system shows that $\tau_{\Delta V1} = 0.9$ for the first and $\tau_{\Delta V2} = 2.0$ for the second regime.

\begin{figure*}
    \centering
    \includegraphics{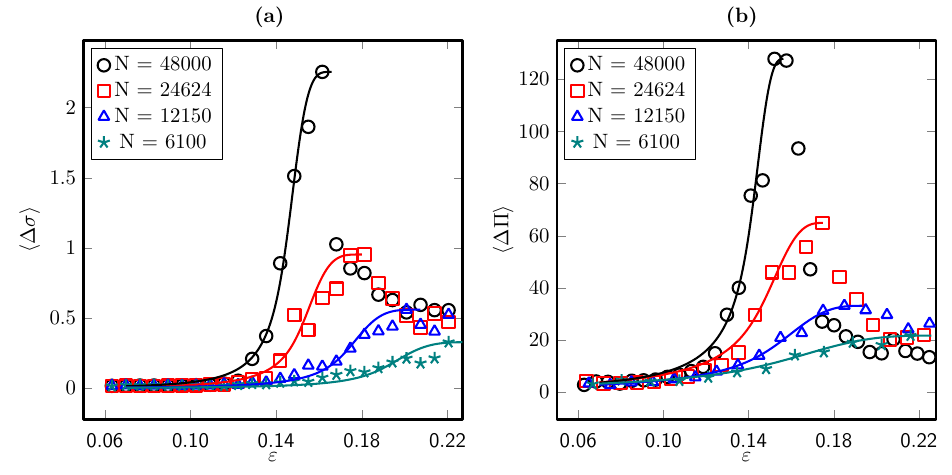}
    \caption{The average avalanche size as a function of the strain for the stress avalanches in (a) and the energy avalanches in (b). The lines represent the best fit for the exponents $\kappa$ and $A$ of the function $(\varepsilon_c-\varepsilon)^{\kappa(\tau-a-1)}\gamma(a-\tau+1, Am_c(\varepsilon_c-\varepsilon)^{\kappa})$ and have been obtained for $\kappa_{\Delta \Pi} = 2.85$ for the energy avalanches and $\kappa_{\Delta \sigma} = 4.0$ for the stress avalanches}
    \label{fig:average_avalanche_per_strain_scaled} 
\end{figure*}

So far, our results were obtained from the PDFs of the events recorded over a strain period from 0 to $\varepsilon_c$. However, the critical exponent $\tau$ from Equation (\ref{eq:fbm_avalanches}) can be obtained by plotting the recorded avalanches at any strain value and over a very small strain range. However, the effect of the exponential cutoff would get stronger as one gets further away from the critical strain \cite{alava}. Therefore, $\tau$ could be evaluated by recording the avalanches as close as possible to the critical strain with a range that is as small as possible while still having enough number of events to represent the statistics. Accordingly, we keep reducing the strain range, until the inclination of the PDF in the log-log plot saturates at $\Delta \varepsilon = 0.005$. We choose the biggest system for this computation since it has the biggest range, over which the lower and upper cutoffs have a small influence. We found that the critical exponent takes a value of $\tau_{\Delta \Pi} \approx 1.4$. From this value, one can compute the exponent $\kappa_{\Delta \Pi} = 1/(\tau'_{\Delta \Pi} - \tau_{\Delta \Pi}) \approx 2.85$. This deviates from the MFT predictions for both the FBM and depinning models which predict $\tau = 1.5$ and $\kappa = 1.0$ \cite{hansen} and $\tau = 1.5$ and $\kappa = 2.0$ \cite{dahmen2009}, respectively. Similarly, by repeating the procedure for the stress drops, one obtains the values $\tau_{\Delta \sigma} = 1.27$ and $\kappa_{\Delta \sigma} = 40$.
By analyzing the critical exponents pertaining to both the energy and stress drops, it is apparent that a significant deviation from the mean-field exponents $\tau = 1.5$ is present. This is not surprising, since it is known that numerous models and simulations do not conform to mean-field calculations. Furthermore, it was shown that damping in sheared glasses has an effect on the critical behavior \cite{salerno}, where over-damped systems have an exponent $\tau = 1.25$ even though the critical behavior is preserved. However, to prove these results, one has to rule out the effect of the initial notch since it favors a localizing effect leading to a concentration of events near the crack tip. Short-range interactions could dominate the system due to the presence of an initial crack, leading to the failure of the mean-field predictions, comparable to the fiber-bundle model with local load-sharing rules if one initially removes some of the fibers.
To investigate this effect we carried out more simulations for the system with 48000 particles but without an initial crack and again computed the critical exponents. The results show that the influence on $\tau_{\Delta \Pi}$ is negligible, which is almost identical to the exponent computed from the notched samples at a value of $1.4$. However, a considerable difference appears in the computation of $\kappa_{\Delta \Pi}$ which has a value of $2.2$. 
As for the stress avalanches, the PDFs are obviously different for the notched specimen. The value of the exponent is $\tau_{\Delta \sigma} = 1.43$, while $\kappa_{\Delta \sigma} = 2.1$. This indicates that the initial notch drives the avalanche exponents of the stress drops away from the MF predictions. This increase in the slope of the avalanche PDF corresponds to the predictions made in the FBM with the local load-sharing rule in the presence of an initial crack \cite{subhadeep}. However, this effect is less pronounced for the exponents of the energy drops, affecting rather $\kappa_{\Delta \Pi}$ than $\tau_{\Delta \Pi}$. Moreover, it seems that the difference between the energy and stress avalanche statistics almost vanishes in the absence of an initial crack. To explore this aspect further, we computed the Pearson correlation coefficient \cite{pearson}, between the energy drops and stress drops and found a consistently high linear correlations $r \approx 0.99$ across the all the systems when considering all the avalanche sizes. However, when considering only the largest avalanche from each sample, we obtain lower values $r \approx 0.93-0.98$ where the smallest system has the lowest linear correlation. Indicating that the effect is stronger for smaller sample sizes due to the larger relative size of the initial crack, which leads to a distortion of the stress avalanche statistics. We conclude that the strength of the disorder in the system, was not enough to suppress the perturbing effect of the initial crack on the distribution of the avalanches over the shape and the size of the simulation cell even though the size was deliberately kept to a minimum while still constituting a critical crack for the failure of the system.

Now we turn our attention to the dependence of the avalanche PDF on the external field, i.e., the strain in our case, by plotting the PDFs for different strain bins. We expect the PDFs to take the following form $P(m, \varepsilon) \sim m^{-\tau-1/k}( \gamma(1/k, Am(\varepsilon_c- \varepsilon)) -  \gamma(1/k, Am(\varepsilon_c))$, where $\tau$, $\kappa$, and $\varepsilon_c$ were computed previously, while A remains unknown.

To investigate this further, the scaling relation between the average avalanche size and the strain is explored. As a first step, we test the fiber bundle relation $\langle m \rangle \sim (\varepsilon_c - \varepsilon)^{\gamma}$. However, we find a disagreement of our data with this relation, where one would expect a linear plot on a double logarithmic scale (at least in the asymptotic limit for large enough events) if the power law were to hold \cite{zapperi1999pre}. We find that the average avalanche size has an exponential relation with the strain, diverging as the systems approach the critical strain. The average avalanche size against the strain for the stress and the energy avalanches is plotted in Figure \ref{fig:average_avalanche_per_strain_scaled}.
It is shown that the average avalanche size diverges between the different sample sizes close to the critical strain. This is qualitatively consistent with Equation (\ref{eq:avg_avalanche_per_strain}) which is derived by analyzing the first moment of the strain-dependent PDF by assuming an exponential cutoff at a critical avalanche size $m_c$. Therefore, we fitted the constant $A$ of this function for the values of the critical exponents $\tau_{\Delta \Pi}$, $\kappa_{\Delta \Pi}$, and $\varepsilon_c$ computed before. For the constant A, we obtained the values $251.5$, $124.7$, $88.7$, and $38.94$ for the systems of sizes $48000$, $24624$, $12150$, and $6100$, respectively.

\begin{figure*}
    \centering
    \includegraphics{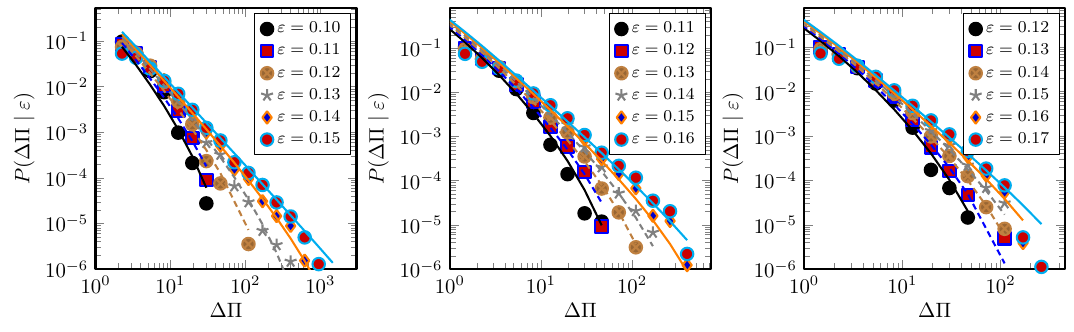}
    \caption{Energy avalanche PDFs for systems with 48000 (a), 24624 (b), and 12150 (c) particles, integrated up to different strain thresholds, represented by the markers. The solid lines are the scaling functions derived from the assumed exponential cutoff $m^{-\tau-1/k}\left( \gamma(1/k, Am(\varepsilon_c- \varepsilon)) -  \gamma(1/k, Am(\varepsilon_c))\right)$ with the universal exponents $\tau_{\Delta \Pi} = 1.4, \kappa_{\Delta \Pi}=2.85$, which was computed from the moment analysis in Equation (\ref{eq:avg_avalanche_per_strain}). The value $\varepsilon_c$ is system-dependent and is shown in Figure \ref{fig:system_size_dependanc}. The parameter $A$ is a system-dependent non-universal constant computed through fitting and shown in Figure \ref{fig:average_avalanche_per_strain_scaled}.}
    \label{fig:energy_pdf_strain_int} 
\end{figure*}

To validate the obtained fits and the derived equations, one can plot the PDFs of the energy avalanches conditioned by the strain. This can be seen in Figure \ref{fig:energy_pdf_strain_int}, where markers represent PDFs that were recorded from 0 to different strain thresholds until reaching the critical strain. The lines represent the derived function with the computed critical exponents and $A$. It is evident that this function gives a reasonable fit for the provided data. 

\section*{Conclusion}

We studied the fracture process of silica glass on the nanoscale and revealed several important findings. We showed that the avalanches follow power law statistics indicating that our systems exhibit critical behavior. However, due to the short period of the steady state before failure, we ruled out a self-organized critical behavior, which is a phenomenon that is usually observed during the plastic yielding of sheared glasses. Furthermore, the results of the computed critical exponents indicate that the approach to failure in the systems without a notch could be described by the depinning models and possibly belong to the same universality class. However, a peculiar behavior is observed for pre-notched systems, where a higher value of the avalanche exponential cutoff exponent $\kappa$ is computed. This increase means that power laws decay faster as the system approaches the critical point since the system fails earlier and large events are less likely to occur. Moreover, the initial notch has a more pronounced effect on the stress avalanche than the energy avalanches. Accordingly, $\tau_{\Delta \Pi}$ demonstrates a minor deviation from the mean-field exponent, while $\tau_{\Delta \sigma}$ diverges significantly from it. This finding indicates that a nucleated crack leads not only to localization of stresses but also perturbs the avalanche statistics. Thus, the avalanches are better studied by analyzing by the dissipated energy. Moreover, the derived relations of the strain-conditioned probability density functions and average avalanche sizes show that the scaling relations, derived from the FBM, fit our results qualitatively. However, mean-field depinning models provides similar predictions with a different exponent $\kappa = 2$ \cite{cyclcic_LJ}, providing further evidence that the they can capture the dynamics of fracture avalanches.

An important issue that should be addressed in future studies is the discrepancy between the computed fractal dimension and the established values. We postulate that this is caused by the initial notch, which leads to smaller critical avalanches. This phenomenon is expected to have a stronger effect in smaller systems, leading to an overestimation of the fractal dimension. However, it is also crucial here to conduct additional investigations to confirm the origin of this inconsistency. 

Furthermore, we made intriguing observations regarding the scaling laws associated with crack growth avalanches. In these observations, different scaling behavior is observed for the small and large avalanche regimes, as shown in Figure \ref{fig:volume_avalanche_pdf_logspace}. Intriguingly, this finding connects with a previous study published by Rountree et al.~\cite{Rountree2007}, in which two different values for the crack roughness for two different scales of silica glass were found. Therefore, further exploring the relationship between crack volume avalanche statistics and crack fractal geometry could provide valuable insights into the underlying mechanisms of crack growth.  

We finally conclude that the observed scaling law indicates the occurrence of a phase transition during the fracture process of silica glass. However, the exact nature of this transition remains unclear at this stage, given that the athermal quasi-static framework, together with finite long-range interactions, could bring our systems to get close to the spinodal point which is associated with power laws of the avalanches similar to a second-order transition. Therefore, further research is needed to determine the order of this transition and potential crossover regions at even larger scales, considering that our systems exhibit increased brittleness on larger scales.


\section*{Appendix}\label{appendix}

\subsection*{Moments analysis}

According to fiber bundle model, the PDF of avalanches bigger than a certain threshold has the following form:

\begin{equation}
    \begin{aligned}
        P(m,\varepsilon) \sim m^{-\tau}f(Am(\varepsilon_c-\varepsilon)^{\kappa})\;, 
    \end{aligned}
    \label{abst_eq:fbm_pdf}
\end{equation}

where $f(Am(\varepsilon_c-\varepsilon)^{\kappa})$ is the upper cutoff function. This can be written as:

\begingroup
\allowdisplaybreaks
\begin{align}
    P(m,\varepsilon) \sim & A^{-1}(\varepsilon_c-\varepsilon)^{\tau\kappa}\left(Am(\varepsilon_c-\varepsilon)^{\kappa})^{-\tau} \right. \nonumber \\ 
    & \left. f(Am(\varepsilon_c-\varepsilon)^{\kappa} \right.)\;, \nonumber \\ \nonumber \\
    \sim & (\varepsilon_c-\varepsilon)^{\tau\kappa}g(Am(\varepsilon_c-\varepsilon)^{\kappa})\;.
\end{align}
\endgroup

As a consequence, the first moment of the PDF can be written as:

\begin{align}
    \langle m^a \rangle &\sim \int_0^{m_c}m^aP(m,\varepsilon)dm\;, \nonumber \\ \nonumber \\
    &\sim \int_0^{m_c}(\varepsilon_c-\varepsilon)^{\tau\kappa}m^ag(Am(\varepsilon_c-\varepsilon)^{\kappa})dm\;.
\end{align}

If we do a variable change: $z = Am(\varepsilon_c-\varepsilon)^{\kappa}$ so $m = A^{-1}z(\varepsilon_c-\varepsilon)^{-\kappa}$ and $dm = A^{-1}(\varepsilon_c-\varepsilon)^{-\kappa}dz$. It follows:

\begin{align}
    \langle m^a \rangle &\sim A^{-1}\int_0^{z_c}(\varepsilon_c-\varepsilon)^{\kappa(\tau-a-1)}z^{a}g(z)dz\;, \nonumber \\ \nonumber \\ 
    &\sim (\varepsilon_c-\varepsilon)^{\kappa(\tau-a-1)}\int_0^{z_c}z^{a}g(z)dz\;.
\end{align}

From the FBM, $f(z)$ is an exponential function $e^{-z}$, so $g(z)$ would be $z^{-\tau}e^{-z}$:

\begin{equation}
    \langle m^a \rangle \sim (\varepsilon_c-\varepsilon)^{\kappa(\tau-a-1)}\int_0^{z_c}z^{a-\tau}e^{-z}dz\;,
\end{equation}

where $z_c = Am_c(\varepsilon_c-\varepsilon)^{\kappa}$. Then, for all values ($a -\tau + 1 > 0$) it follows:

\begin{equation}
    \int_0^{z_c}z^{a-\tau}e^{-z}dz = \gamma(a-\tau+1, Am_c(\varepsilon_c-\varepsilon)^{\kappa})\;,
\end{equation}

where $\gamma(a-\tau+1, Am_c(\varepsilon_c-\varepsilon)^{\kappa})$ is the lower incomplete gamma function. So the first moment of the PDF is written as:

\begin{equation}
    \langle m^a \rangle \sim (\varepsilon_c-\varepsilon)^{\kappa(\tau-a-1)}\gamma(a-\tau+1, Am_c(\varepsilon_c-\varepsilon)^{\kappa})\;.
\end{equation}

\subsection*{Strain dependence}

Starting from Equation (\ref{abst_eq:fbm_pdf}) and integrating over the strain range 0 to $\varepsilon_c$:
\begingroup
\allowdisplaybreaks
\begin{align}  
    P(m,\varepsilon) &\sim \int_0^{\varepsilon_c}  m^{-\tau}f(Am(\varepsilon_c-\varepsilon)^{\kappa})d\varepsilon\;, \nonumber \\ \nonumber \\
    &\sim  m^{-\tau} \int_0^{\varepsilon_c} e^{-Am(\varepsilon_c-\varepsilon)^{\kappa}}d\varepsilon\;, \nonumber \\ \nonumber \\
    &\sim  m^{-\tau} \int_{m\varepsilon_c^{\kappa}}^0 (\kappa^{-1}A^{-1} m^{-1/\kappa}z^{1/\kappa - 1}e^{-z}dz)\;, \nonumber \\ \nonumber \\
    &\sim  \kappa^{-1}A^{-1} m^{-\tau}m^{-1/\kappa} \int_{m\varepsilon_c^{\kappa}}^0 z^{1/\kappa - 1} e^{-z}dz\;, \nonumber \\ \nonumber \\
    &\sim  m^{-\tau-1/\kappa} \int_{m\varepsilon_c^{\kappa}}^0 z^{1/\kappa - 1} e^{-z}dz\;, \nonumber \\ \nonumber \\
    &\sim  m^{-\tau-1/\kappa} \gamma(1/\kappa, m\varepsilon_c^k)\;. 
\end{align}
\endgroup

Similarly, for arbitrary bin sizes $\varepsilon_2 - \varepsilon_1$:
\begingroup
\allowdisplaybreaks
\begin{align}
    P(m,\varepsilon) \sim  m^{-\tau-1/\kappa} &\int_{Am(\varepsilon_c-\varepsilon_1)^{\kappa}}^{Am(\varepsilon_c-\varepsilon_2)^{\kappa}} z^{1/\kappa - 1} e^{-z}dz\;, \nonumber \\ \nonumber \\
    \sim  m^{-\tau-1/\kappa} &\left(\int_{0}^{Am(\varepsilon_c-\varepsilon_2)^{\kappa}} z^{1/\kappa - 1} e^{-z}dz - \right. \nonumber \\ \nonumber \\
    &\left. \int_{0}^{Am^l(\varepsilon_c-\varepsilon_1)^{\kappa}} z^{1/\kappa - 1} e^{-z}dz  \right)\;, \nonumber \\ \nonumber \\
   \sim m^{-\tau-l/\kappa}  &\left( \gamma(1/\kappa, Am(\varepsilon_c-\varepsilon_2)^{\kappa}) - \right. \nonumber \\ \nonumber \\
    &\left. \gamma(1/\kappa, Am(\varepsilon_c-\varepsilon_1)^{\kappa}) \right)\;.
\end{align}
\endgroup

Finally, if $\varepsilon_1 = 0 $, the PDF is written as:

\begin{equation}
    \begin{split}
    P(m, \varepsilon) \sim  m^{-\tau-1/\kappa} \left( \gamma(1/\kappa, Am(\varepsilon_c-\varepsilon)^{\kappa}) - \right.\\ \left. \gamma(1/\kappa, Am(\varepsilon_c)^{\kappa}) \right)\;.
    \end{split}
\end{equation}

\end{document}